# Controlled fabrication of freestanding monolayer SiC by electron irradiation

Yunli Da (笪蕴力)[1,#], Ruichun Luo (罗瑞春)[1,#], Bao Lei (雷宝)[1],

Wei Ji (季威)[2], Wu Zhou (周武)[1,*]

[1]School of Physical Sciences, University of Chinese Academy of Sciences, Beijing 100049, China

[2]Department of Physics and Beijing Key Laboratory of Optoelectronic Functional Materials & Micro-Nano Devices, Renmin University of China, Beijing 100872, China

[#] These authors contributed equally to this work.

*Corresponding author: wuzhou@ucas.ac.cn (W.Z.)

## Abstract

The design and preparation of novel quantum materials with atomic precision are crucial for exploring new physics and for device applications. Electron irradiation has demonstrated as an effective method for preparing novel quantum materials and quantum structures that could be challenging to obtain otherwise. It features the advantages of precise control over the patterning of such new materials and their integration with other materials with different functionalities. Here, we present a new strategy for fabricating freestanding monolayer SiC within nanopores of a graphene membrane. By regulating the energy of the incident electron beam and the in-situ heating temperature in a scanning transmission electron microscope (STEM), we can effectively control the patterning of nanopores and subsequent growth of monolayer SiC within the graphene lattice. The resultant SiC monolayers seamlessly connect with the graphene lattice, forming a planar structure distinct by a wide direct bandgap. Our in-situ STEM observations further uncover that the growth of monolayer SiC within the graphene nanopore is driven by a combination of bond rotation and atom extrusion, providing new insights into the atom-by-atom self-assembly of freestanding two-dimensional (2D) monolayers.

1. Introduction

2D materials exhibit extraordinary properties that often outperform those of their bulk counterparts. This has been substantiated through extensive research on a variety of layered materials, including graphene[1-3] and transition metal dichalcogenides (TMDs).[4-6] Controlled fabrication of such 2D materials into monolayers have been well explored by both top-down exfoliation as facilitated by the van der Waals interactions between adjacent layers, and bottom-up growth techniques.[7-10] However, synthesizing monolayers from non-layered bulk materials poses a significant challenge due to the robust chemical bonds among their constituent atoms, not to mention the additional complexity of regulating the morphology and composition.[11] A prime example is bulk SiC, a non-layered indirect wide-bandgap semiconductor extensively utilized across diverse technological domains.[12,13] Theoretical predictions suggest that monolayer SiC is thermodynamically stable and exhibits a direct bandgap.[14,15] Despite several experimental attempts,[16-19] the successful synthesis of freestanding monolayer SiC remains elusive. The studies reported to date have been limited to the formation of a SiC nanoseed containing 6 Si atoms within a graphene lattice[18] or epitaxial growth of monolayer SiC on a TaC substrate.[19] The latter approach, in particular, has resulted in strong electronic coupling with the substrate, complicating the exploration of the intrinsic characteristics of freestanding monolayer SiC.

On the other hand, electron microscopy has proven indispensable for investigating the intricate local structures and properties of suspended monolayer 2D materials.[20-22] Meanwhile, the interaction with an energetic electron beam (e-beam) can sometimes induce structural fluctuations and modifications within the observed samples due to energy transfer, which can be subsequently harnessed for materials processing.[23-25] In particular, in aberration-corrected STEM, an atomic-sized focused electron probe can enable atomic-scale structural manipulation, in some cases even with single-atom precision.[26-30] As a result, the focused STEM probe can be used to facilitate atomic-level structural destruction,[31,32] reconstruction,[33,34] defect healing,[35,36] and controlled fabrication of novel monolayer materials and quantum structures[37-41]. For example, controlled electron irradiation has been applied to fabricate monolayers of Fe [37] and CuO[40] by inducing self-assembly of impurity atoms on graphene into novel monolayer structures. In addition, monolayer Mo[41] and metallic MX (M=Mo, W; X= S, Se) nanowires of three-atoms wide[38,39] have been sculpted out of 2D monolayers of semiconducting TMDs by inducing chalcogen vacancies and subsequent driving the structures into their thermodynamically stable forms. The STEM instruments offer the flexibility of fine adjusting the energy and dose rate of the incident e-beam to regulate the energy transfer to the samples. In addition, a variety of external fields can be applied via in-situ holders to facilitate in-situ (S)TEM studies under different stimuli.[42-45] Among them, in-situ heating has been frequently employed to observe thermally induced structural transformations,[42,43] as well as growth mechanisms at the atomic scale. These advancements not only offer new insights into the structure-property relationship of 2D materials, but also pave the way for innovative approaches to materials engineering and manufacturing.

In this work, we report a novel workflow to fabricate freestanding monolayer SiC confined within the graphene lattice using an aberration-corrected STEM. The workflow involves accurate control of the e-beam energy and in-situ heating temperature, allowing for selective sculpting of graphene nanopores and directional deposition and assembly of SiC at different stages. Through quantitative STEM analysis and density-functional theory (DFT) simulations, we demonstrate that the freestanding monolayer SiC embedded in graphene nanopores adopts a stable planar structure with a direct bandgap of 2.56 eV. In-situ STEM imaging reveals that the atomic-scale growth of monolayer SiC is driven by bond rotation and atom extrusion, synergistically promoting the atom-by-atom assembly of the SiC lattice. Our results demonstrate the feasibility of fabricating monolayer SiC in freestanding form, and highlight the potential for e-beam assisted large-area programmable patterning of novel 2D materials for potential applications in electronic devices.

## 2. Results and discussion

The controlled growth of freestanding monolayer SiC was performed inside a STEM during in-situ heating. As illustrated in Figure 1a, the entire process involves four critical steps in sequence: 1) cleaning of the graphene surface, 2) sculpting of graphene nanopores, 3) deposition of Si sources, and 4) self-assembly of SiC monolayer from diffused Si and C atoms. The graphene sample was synthesized using a chemical vapor deposition (CVD) method and transferred onto customized micro-electro-mechanical-system (MEMS) chips through a polymethyl methacrylate (PMMA) assisted wet-transfer process (details in the *Methods*). During sample growth and transfer, residual contaminants often accumulate on the graphene surface, primarily consisting of Si, C, O and H elements (Figure S1). These amorphous residues are undesirable for atomic-scale analysis of the graphene sample, and would severely interfere with the fabrication of monolayer SiC by hindering the surface diffusion of Si and C atoms. To address this contamination issue, we applied in-situ heating at 550°C inside the UHV environment of the STEM (see methods) to remove contaminants from the as-prepared graphene, with the e-beam blanked during this step. As shown in Figure 1b, the in-situ heating at 550°C can effectively remove most of the hydrocarbon contaminants from the graphene surface, as demonstrated by the uniform STEM annular dark field (ADF) image contrast for both monolayer and bilayer regions and the clear atomic resolution STEM-ADF image of the graphene lattice shown in Figure 1f. Only some minor residues, primarily containing Si impurities, can be observed after in-situ heating, concentrating along grain boundaries and step edges of the graphene sample.

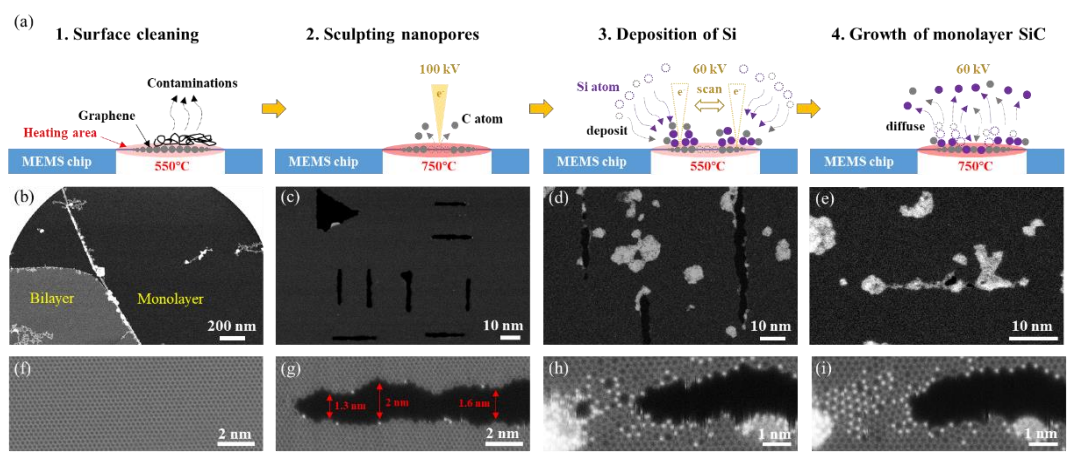

**Fig. 1 E-beam assisted growth of freestanding monolayer SiC.** (a) Schematic of the four steps for controlled growth of monolayer SiC in graphene nanopores by using a STEM. (b-e) Corresponding low-magnification STEM-ADF images of the graphene sample at the four different steps. (f-i) Corresponding atomic-scale STEM-ADF images of the sample at the four different steps as shown in (a).

In the second step, we demonstrate the capability to sculpt graphene nanopores at designated locations and in various shapes with sub-nm precision using a high energy focused e-beam. It has been demonstrated that the knock-on damage threshold of graphene is approximately 80 kV.[46] Below 80 kV, the pristine lattice of graphene can be imaged under high vacuum with a high electron dose and still maintains its structural integrity. In contrast, above 80 kV, e.g., at 100 kV, knock-on damage will cause continuous sputtering of carbon atoms from the graphene lattice. By utilizing the knock-on damage effect at 100 kV and the capability to precisely control the scanning of the e-beam, we successfully sculpt nanopores in monolayer graphene with various shapes, dimensions and orientations, as shown in Figure 1c. To accelerate the sputtering process and to prevent re-deposition of the knocked-out carbon atoms, we performed this sculpting process under 100 kV and at 750 °C. Under our experimental conditions, a 30 nm long and 4 nm wide rectangular nanopore can be created within the graphene lattice by continuous scanning of the e-beam along predefined paths for about 5 minutes. A time sequence of STEM-ADF imaging of the sculpting process can be seen in Figure S2. Figure 1g shows an as-prepared graphene nanopore along the zigzag direction, with a width of less than 2 nm. The edges of the nanopore are somewhat irregular, terminated by bonding unsaturated C atoms and occasionally decorated by isolated Si atoms (Figure 1g). The efficient sculpting process at 100 kV and 750 °C demonstrates the potential for programmable patterning of graphene nanopores in STEM, albeit further refinement of the experimental parameters, with the capability to create scalable integration patterns for nanodevices.

In order to grow monolayer SiC into the graphene nanopores, we need additional Si and C sources around the nanopores. As shown in the schematics in Figure 1a, during the third step, the heating temperature was lowered to 550°C, and the acceleration voltage of the STEM was reduced to 60 kV. This adjustment allowed for non-destructive scanning and imaging of the pristine graphene lattice, while slight rearrangement of carbon atoms along the graphene nanopore edges still occurs under the e-beam. It was noted that under this particular experimental setting (60 kV, 550°C), the e-beam illumination would induce deposition of Si and C atoms onto the sample, and these deposits can aggregate into nanoclusters showing higher STEM-ADF contrast (Figure 1d). Electron energy-loss spectroscopy (EELS) measurement confirms that the nanoclusters are composed of Si and C without oxygen (Figure S3). The nanoclusters could further grow with prolonged scanning of the e-beam (Figure S4), indicating a steady supply of Si and C from the environment. As we used customized $SiN_x$ membranes with penetrated holes to support the graphene sample, it is likely that the extra Si atoms are thermally activated from the amorphous $SiN_x$. A closer look at the nanopore region after Si deposition (Figure 1h) revealed that many isolated Si atoms had doped into the graphene lattice or anchored at the nanopore edges, surrounded by disordered C rings. By precisely controlling the e-beam energy and heating temperature, we achieved smooth deposition of Si-C nanoclusters, which can serve as source materials for subsequent self-assembly of monolayer SiC.

In the fourth step, we maintained the e-beam energy at 60 kV to avoid knock-on damage, but increased the heating temperature to 750 °C to facilitate self-assembly of Si and C atoms into the graphene nanopores to form monolayer SiC. During this process, the e-beam was blanked to avoid interfering the thermal-driven process, and was only used for imaging and analysis of the resultant structure. As shown in Figure 1e, the 2-nm-wide nanopore slot is filled with atoms showing higher contrast than graphene but lower contrast than that of the $SiC_x$ nanoclusters. Figure 1i depicts the atomic resolution STEM-ADF image of the same nanopore shown in Figure 1h. A comparison between these two images suggests that the randomly distributed Si atoms within the disordered carbon lattice in the third step have transformed into a more regular arrangement with neighboring carbon atoms, indicating successful self-assembly of freestanding monolayer SiC embedded within the graphene lattice. The growth of SiC is likely due to the feeding from adjacent $SiC_x$ nanoclusters, as evidenced by their presence in the left corners of Figure 1h and 1i. The continuous filling of Si atoms and the crystallization of SiC monolayer are facilitated by the elevated temperature during in-situ heating in the STEM. It is worth noting that the nanopore in Figure 1i is not completely filled with monolayer SiC, leaving a 1.5 nm wide gap on the right-hand-side. This may be caused by the limited heating time (about 1h) or insufficient source supply.

Following the same methodology, another nanopore slot, less than 2 nm wide, was sculpted and filled with monolayer SiC by maintaining the last step at 750°C for 6 hours. Figure 2a shows the atomic structure of the as-grown monolayer SiC domains. Notably, the monolayer SiC domains are not seamlessly connected but instead separated by disordered carbon rings, and two unfilled nanopores of approximately 1 x 2 nm can also be observed. This imperfection may result from the competing growth of monolayer SiC and amorphous carbon at 750 °C, suggesting that further refinements of the experimental procedure and parameters are still needed in order to grow freestanding monolayer SiC into larger sizes.

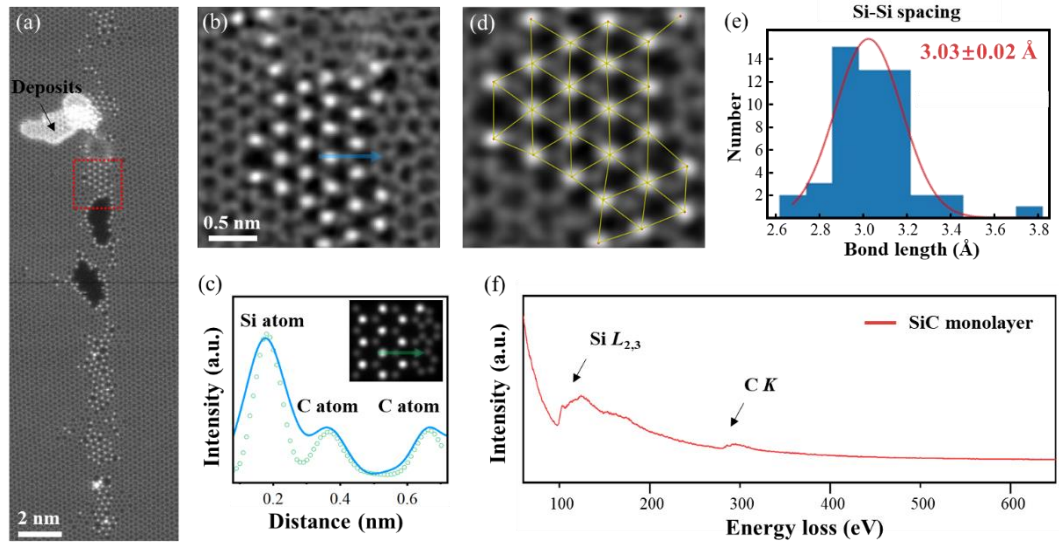

**Fig. 2 Structural and chemical analysis of monolayer SiC.** (a) STEM-ADF image showing the structure of the as-grown monolayer SiC within the graphene lattice. (b) The zoomed-in image from the red dashed box in (a). (c) Intensity line profile showing the STEM-ADF contrast of Si and C atoms. The blue curve is from the selected blue region in (b); The green circles are the data points for simulated results. The inset shows the simulated STEM-ADF image from the monolayer SiC-graphene structural model. (d) Illustration of calculating the Si-Si distance in the SiC lattice from the experimental STEM-ADF image in (b). (e) Statistics of the Si-Si distance in (d). (f) EELS of a monolayer SiC domain, confirming its composition of only Si and C.

We now turn to analyze the atomic structure of the as-formed monolayer SiC. Figure 2b depicts a typical monolayer SiC nanocrystal of 1.2 x 2.1 nm. The filtered STEM-ADF image, magnified from the red box in Figure 2a, clearly reveals a hexagonal lattice structure, characteristic of ordered Si-C bonding. STEM-ADF imaging, known for its atomic number (Z) contrast[47,48], is now well recognized as an effective method for atom-by-atom chemical analysis of monolayer materials[49,50]. The intensity line profile along the blue box in the experimental STEM-ADF image, as depicted in Figure 2c, indicates that the lighter atoms in the Si-C lattice are single carbon atoms, as their intensity matches that of the carbon atoms in the graphene lattice. Furthermore, as compared with the simulated STEM-ADF image using the monolayer SiC-graphene structural model (the inset in Figure 2c), the brighter atoms can be identified as single Si atoms. The SiC structure was further verified by EELS analysis, showing the presence of only Si and C (Figure 2f). The STEM-ADF imaging and EELS analyses conclusively demonstrate the successful formation of freestanding monolayer SiC crystal embedded in the graphene lattice.

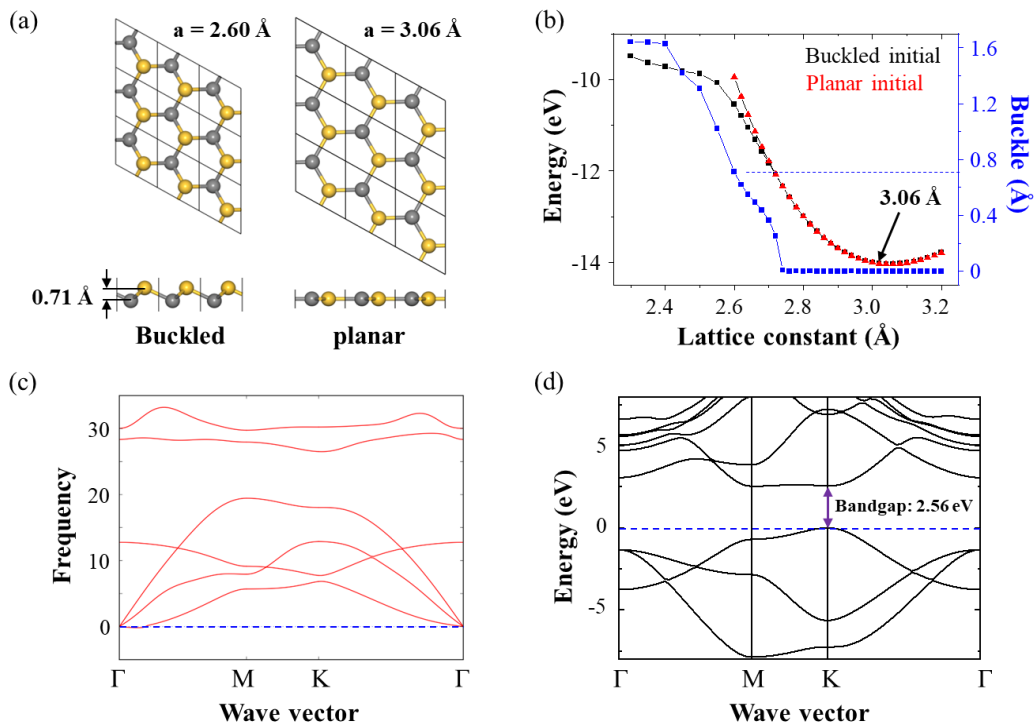

**Fig. 3 Atomic structure and physical properties of freestanding monolayer SiC.** (a) Structure models of SiC with either buckled or planar configurations. (b) Calculated

total energy and buckling of a monolayer SiC supercell as a function of the lattice constant. (c) The phonon diagram of freestanding monolayer SiC in planar configuration. (d) The electronic structure of freestanding monolayer SiC in planar configuration.

Note that STEM-ADF images are 2D projections of the sample, which could not directly reveal the full 3D structures based on a single image except for in some special cases[51]. Bulk SiC is a non-layered material, in which the Si and C atoms adopt the $sp^3$ hybridization, i.e., the Si and C atoms are not in the same atomic plane in bulk SiC. If monolayer SiC retains its bulk characteristics, it would exhibit a buckled structure with an in-plane Si-Si projected spacing of 2.6 Å, as illustrated in Figure 3a. However, our measured Si-Si projected spacing is 3.03 Å in the as-grown freestanding monolayer SiC (Figure 2d and 2e), significantly larger than that of the buckled SiC structure. This observation aligns with a previous theoretical prediction that monolayer SiC could adopt a planar configuration where all the Si and C atoms are in the same atomic layer, forming the $sp^2$ hybridization[15]. Our DFT calculations indicate the buckled C-Si bond becomes fully flattened at a lattice constant of 2.74 Å, forming a planar monolayer (Figure 3b). The equilibrium lattice constant, namely the Si-Si spacing of this monolayer was predicted to be 3.06 Å (Figure 3b), highly consistent with our experimentally observed 3.03 Å. This consistency supports that the proposed planar structure shown in Figure 3a represents the experimentally observed structure. We further calculated the phonon dispersion spectra (Figure 3c) of this freestanding planar monolayer SiC. No imaginary frequency was observed in the spectra, suggesting the planar SiC monolayer is dynamically stable. The electronic band structure (Figure 3d) reveals that the planar SiC monolayer is a direct bandgap semiconductor with a wide bandgap of 2.56 eV.

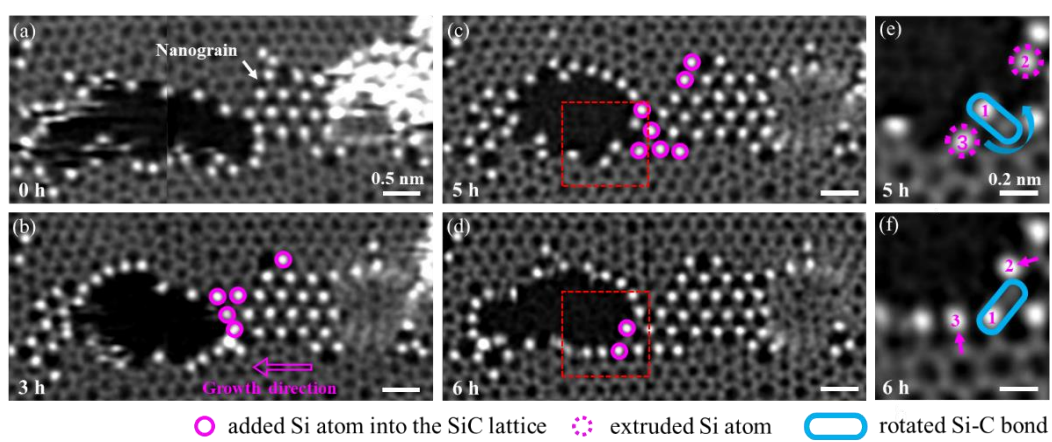

**Fig. 4 Structural evolution during the in-situ growth of monolayer SiC.** (a-d) Time sequence STEM-ADF images showing the atomic-scale structures around the SiC lattice collected at 0 h, 3 h, 5 h and 6 h, respectively, during the in-situ annealing at 750 °C, showing the growing of the monolayer SiC domain. The 0 h initial state is set arbitrarily during a prolonged annealing process. (e, f) The zoom-in STEM-ADF images from the red dashed boxes in (c) and (d), respectively.

To further unveil the growth mechanism of freestanding monolayer SiC, we conducted time sequence STEM-ADF imaging at 60kV during the growth of the monolayer SiC domains. The sequential STEM-ADF images are shown in Figure 4a-d, revealing the structural evolution of the monolayer SiC nanocrystal. Initially, a small SiC nanograin was observed (Figure 4a). As the annealing time at 750°C increased, additional Si and

C atoms gradually diffused from nearby SiC$_x$ nanoclusters. The magenta solid circles highlight the newly added Si atoms as compared to the previous snapshot, suggesting the atom-by-atom growth of the monolayer SiC nanocrystal along the specified direction. Moreover, Si atoms tend to anchor at the edges of the nanopores,[52] which can guide the atom migration and facilitate the reconstruction of the intermediate structures into a regular SiC lattice. Comparison of the two snapshots in Figure 4c and Figure 4d indicates the addition of only two Si-C units into the SiC nanodomain, which provides an excellent chance to elucidate the atomic-scale transformation pathways for the growth of monolayer SiC domains. As shown in Figures 4e-f, the Si atom labeled "1" moved to the adjacent C atom position via a Si-C bond rotation, as highlighted by the blue markings. In addition, the edge Si atoms could be extruded by the incorporation of C atoms into their original positions, as illustrated by the Si atoms in magenta dashed circles (labeled 2 and 3). The Si-C bond rotation and atom extrusion due to C incorporation work in concert to promote the growth of monolayer SiC, driven by atomic diffusion and thermodynamics at high annealing temperatures.

## 3. Conclusion

In summary, we have designed a patternable workflow in an aberration-corrected STEM to intentionally grow freestanding monolayer SiC within the graphene lattice. By precisely controlling the e-beam energy and heating temperature, we have demonstrated the selective sculpture of graphene nanopores and the directional growth of monolayer SiC domains into the nanopores while preserving the integrity of the surrounding graphene structures. Combining STEM analysis and DFT calculations, we validated the stable planar structure of freestanding monolayer SiC, and unveiled that such planar monolayer SiC structure possesses a direct bandgap of 2.56 eV. Through in-STEM observation, we identified the synergistic processes of bond rotation and atom extrusion that promote the growth of the monolayer SiC lattice. We acknowledge that the obtained freestanding monolayer SiC still has a relatively small size, but we expect that our method could in principle be further optimized to fabricate large-scale monolayer SiC as this structure is thermodynamically favorable. Our study paves the way for atomic-scale manufacturing of novel quantum materials with large-area programmable patterning using the electron beam.

**Methods**

**Sample preparation.** Monolayer graphene was grown on a copper film using a CVD system. The copper film was firstly cleaned and polished to prepare a flat surface. Then it was put into the center of the tube in the furnace. After flushing with Ar, the copper film was annealed at 1070 °C with Ar/$H_2$ mixed gas ($H_2$: 10% vt) to reduce the oxidized surface at 100 sccm. After 60 min, Ar/$CH_4$ mixed gas ($CH_4$: 0.1% vt) was introduced as the carbon source to grow graphene. The growth time is about 20 min. The as-grown graphene sample on the copper film was transferred onto the MEMS chip by a PMMA-assisted method. The customized MEMS chip with amorphous $SiN_x$ thin film as the window material was used for in-situ heating. Perforated regions were created by FIB etching of the $SiN_x$ film.

**Electron microscopy.** The experiments were performed on a Nion U-HERMES microscope operated at 60 kV or 100 kV under near-ultrahigh vacuum (~2 × 10$^{-7}$ Pa). The beam current for imaging during the experiments was set to 18 pA. The convergence semi-angle for the electron beam was 32 mrad and the semi-angular range of the ADF detector was 44–210 mrad. The in-situ heating was controlled via the customized MEMS chips.

**Density functional theory (DFT) calculations.** The calculations were carried out using projector-augmented wave approach (PAW)[53] and the Perdew-Burke-Ernzerhof (PBE) developed exchange-correlation functional,[54] as implemented in the VASP.[55] The cutoff energy was chosen at 500 eV, and the Brillouin zone was sampled using k-points of 11 × 11 × 1. The convergence thresholds for energy and atomic forces were set as 1 × 10$^{-6}$ eV and 0.01 eV/Å, respectively. The distance of vacuum space was set to larger than 20 Å. The phonon spectrum was computed with density functional perturbation theory (DFPT)[56] and post-treated by Phonopy code.[57]

**Acknowledgments**

This research benefited from resources and supports from the Electron Microscopy Center at the University of Chinese Academy of Sciences. This research is financially supported by the Ministry of Science and Technology (MOST) of China (Grant No. 2018YFE0202700), the Beijing Outstanding Young Scientist Program (BJJWZYJH01201914430039), the China National Postdoctoral Program for Innovative Talents (BX2021301), the Fundamental Research Funds for the Central

Universities, and the Research Funds of Renmin University of China (Grants No. 22XNKJ30). Calculations were performed at the Physics Lab of High-Performance Computing (PLHPC) and the Public Computing Cloud (PCC) of Renmin University of China.

**References**

[1]   Novoselov K S, Geim A K, Morozov S V, Jiang D, Zhang Y, Dubonos S V, Grigorieva I V and Firsov A A. Electric field effect in atomically thin carbon films. ***Science***, 306, 666-669 (2004).

[2]   Novoselov K S, Geim A K, Morozov S V, Jiang D, Katsnelson M I, Grigorieva I V, Dubonos S V and Firsov A A. Two-dimensional gas of massless Dirac fermions in graphene. ***Nature***, 438, 197-200 (2005).

[3]   Castro Neto A H, Guinea F, Peres N M R, Novoselov K S and Geim A K. The electronic properties of graphene. ***Rev. Mod. Phys.***, 81, 109-162 (2009).

[4]   Mak K F, Lee C, Hone J, Shan J and Heinz T F. Atomically Thin $MoS_2$: A New Direct-Gap Semiconductor. ***Phys. Rev. Lett.***, 105, 136805 (2010).

[5]   Wang Q H, Kalantar-Zadeh K, Kis A, Coleman J N and Strano M S. Electronics and optoelectronics of two-dimensional transition metal dichalcogenides. ***Nat. Nanotechnol.***, 7, 699-712 (2012).

[6]   Manzeli S, Ovchinnikov D, Pasquier D, Yazyev O V and Kis A. 2D transition metal dichalcogenides. ***Nat. Rev. Mater.***, 2, 1-15 (2017).

[7]   Pan Y, Shi D-X and Gao H-J. Formation of graphene on Ru (0001) surface. ***Chin. Phys.***, 16, 3151-3153 (2007).

[8]   Pan Y, Zhang H, Shi D, Sun J, Du S, Liu F and Gao H-j. Highly Ordered, Millimeter-Scale, Continuous, Single-Crystalline Graphene Monolayer Formed on Ru (0001). ***Adv. Mater.***, 21, 2777 (2009).

[9]   Li X, Cai W, An J, Kim S, Nah J, Yang D, Piner R, Velamakanni A, Jung I, Tutuc E, Banerjee S K, Colombo L and Ruoff R S. Large-Area Synthesis of High-Quality and Uniform Graphene Films on Copper Foils. ***Science***, 324, 1312-1314 (2009).

[10]  Najmaei S, Liu Z, Zhou W, Zou X, Shi G, Lei S, Yakobson B I, Idrobo J-C, Ajayan P M and Lou J. Vapour phase growth and grain boundary structure of molybdenum disulphide atomic layers. ***Nat. Mater.***, 12, 754-759 (2013).

[11]  Zhou N, Yang R and Zhai T. Two-dimensional non-layered materials. ***Mater. Today Nano***, 8, 100051 (2019).

[12]  Casady J B and Johnson R W. Status of silicon carbide (SiC) as a wide-bandgap semiconductor for high-temperature applications: A review. ***Solid-State Electron.***, 39, 1409-1422 (1996).

[13] Wu R, Zhou K, Yue C Y, Wei J and Pan Y. Recent progress in synthesis, properties and potential applications of SiC nanomaterials. *Prog. Mater Sci.*, 72, 1-60 (2015).

[14] Shi Z, Zhang Z, Kutana A and Yakobson B I. Predicting Two-Dimensional Silicon Carbide Mono layers. *ACS Nano*, 9, 9802-9809 (2015).

[15] Chabi S and Kadel K. Two-Dimensional Silicon Carbide: Emerging Direct Band Gap Semiconductor. *Nanomaterials*, 10, 2226 (2020).

[16] Lin S S. Light-Emitting Two-Dimensional Ultrathin Silicon Carbide. *J. Phys. Chem. C*, 116, 3951-3955 (2012).

[17] Lin S, Zhang S, Li X, Xu W, Pi X, Liu X, Wang F, Wu H and Chen H. Quasi-Two-Dimensional SiC and SiC$_2$: Interaction of Silicon and Carbon at Atomic Thin Lattice Plane. *J. Phys. Chem. C*, 119, 19772-19779 (2015).

[18] Susi T, Skákalová V, Mittelberger A, Kotrusz P, Hulman M, Pennycook T J, Mangler C, Kotakoski J and Meyer J C. Computational insights and the observation of SiC nanograin assembly: towards 2D silicon carbide. *Sci. Rep.*, 7, 4399 (2017).

[19] Polley C, Fedderwitz H, Balasubramanian T, Zakharov A, Yakimova R, Bäcke O, Ekman J, Dash S, Kubatkin S and Lara-Avila S. Bottom-up growth of monolayer honeycomb SiC. *Phys. Rev. Lett.*, 130, 076203 (2023).

[20] Zhou W, Zou X, Najmaei S, Liu Z, Shi Y, Kong J, Lou J, Ajayan P M, Yakobson B I and Idrobo J-C. Intrinsic Structural Defects in Monolayer Molybdenum Disulfide. *Nano Lett.*, 13, 2615-2622 (2013).

[21] Xu M, Bao D-L, Li A, Gao M, Meng D, Li A, Du S, Su G, Pennycook S J, Pantelides S T and Zhou W. Single-atom vibrational spectroscopy with chemical-bonding sensitivity. *Nat. Mater.*, 22, 612-618 (2023).

[22] Xu M, Li A, Pennycook S J, Gao S-P and Zhou W. Probing a Defect-Site-Specific Electronic Orbital in Graphene with Single-Atom Sensitivity. *Phys. Rev. Lett.*, 131, 186202-186202 (2023).

[23] Gonzalez-Martinez I G, Bachmatiuk A, Bezugly V, Kunstmann J, Gemming T, Liu Z, Cuniberti G and Ruemmeli M H. Electron-beam induced synthesis of nanostructures: a review. *Nanoscale*, 8, 11340-11362 (2016).

[24] Zhao X, Kotakoski J, Meyer J C, Sutter E, Sutter P, Krasheninnikov A V, Kaiser U and Zhou W. Engineering and modifying two-dimensional materials by electron beams. *MRS Bull.*, 42, 667-676 (2017).

[25] Luo R, Gao M, Wang C, Zhu J, Guzman R and Zhou W. Probing functional structures, defects, and interfaces of 2D transition metal dichalcogenides by electron microscopy. *Adv. Funct. Mater.*, 34, 2307625 (2024).

[26] Ishikawa R, Mishra R, Lupini A R, Findlay S D, Taniguchi T, Pantelides S T and Pennycook S J. Direct Observation of Dopant Atom Diffusion in a Bulk Semiconductor Crystal Enhanced by a Large Size Mismatch. *Phys. Rev. Lett.*, 113, 155501 (2014).

[27] Tripathi M, Mittelberger A, Pike N A, Mangler C, Meyer J C, Verstraete M J, Kotakoski J and Susi T. Electron-Beam Manipulation of Silicon Dopants in Graphene. *Nano Lett.*, 18, 5319-5323 (2018).

[28] Yang S-Z, Sun W, Zhang Y-Y, Gong Y, Oxley M P, Lupini A R, Ajayan P M, Chisholm M F, Pantelides S T and Zhou W. Direct Cation Exchange in Monolayer MoS$_2$ via Recombination-Enhanced Migration. **Phys. Rev. Lett.**, 122, 106101 (2019).

[29] Su C, Tripathi M, Yan Q-B, Wang Z, Zhang Z, Hofer C, Wang H, Basile L, Su G, Dong M, Meyer J C, Kotakoski J, Kong J, Idrobo J-C, Susi T and Li J. Engineering single-atom dynamics with electron irradiation. **Sci. Adv.**, 5, eaav2252 (2019).

[30] Dyck O, Ziatdinov M, Lingerfelt D B, Unocic R R, Hudak B M, Lupini A R, Jesse S and Kalinin S V. Atom-by-atom fabrication with electron beams. **Nat. Rev. Mater.**, 4, 497-507 (2019).

[31] Lehtinen O, Kurasch S, Krasheninnikov A and Kaiser U. Atomic scale study of the life cycle of a dislocation in graphene from birth to annihilation. **Nat. Commun.**, 4, 2098 (2013).

[32] Pan Y, Lei B, Qiao J, Hu Z, Zhou W and Ji W. Selective linear etching of monolayer black phosphorus using electron beams. **Chin. Phys. B**, 29, 086801 (2020).

[33] Lin Y-C, Dumcenco D O, Huang Y-S and Suenaga K. Atomic mechanism of the semiconducting-to-metallic phase transition in single-layered MoS$_2$. **Nat. Nanotechnol.**, 9, 391-396 (2014).

[34] Lin J, Pantelides S T and Zhou W. Vacancy-Induced Formation and Growth of Inversion Domains in Transition-Metal Dichalcogenide Monolayer. **ACS Nano**, 9, 5189-5197 (2015).

[35] Hopkinson D G, Zolyomi V, Rooney A P, Clark N, Terry D J, Hamer M, Lewis D J, Allen C S, Kirkland A I, Andreev Y, Kudrynskyi Z, Kovalyuk Z, Patane A, Fal'ko V I, Gorbachev R and Haigh S J. Formation and Healing of Defects in Atomically Thin GaSe and InSe. **ACS Nano**, 13, 5112-5123 (2019).

[36] Zhao X, Ji Y, Chen J, Fu W, Dan J, Liu Y, Pennycook S J, Zhou W and Loh K P. Healing of Planar Defects in 2D Materials via Grain Boundary Sliding. **Adv. Mater.**, 31, 1900237 (2019).

[37] Zhao J, Deng Q, Bachmatiuk A, Sandeep G, Popov A, Eckert J and Rümmeli M H. Free-standing single-atom-thick iron membranes suspended in graphene pores. **Science**, 343, 1228-1232 (2014).

[38] Lin J, Cretu O, Zhou W, Suenaga K, Prasai D, Bolotin K I, Cuong N T, Otani M, Okada S and Lupini A R. Flexible metallic nanowires with self-adaptive contacts to semiconducting transition-metal dichalcogenide monolayers. **Nat. Nanotechnol.**, 9, 436-442 (2014).

[39] Lin J, Zhang Y, Zhou W and Pantelides S T. Structural Flexibility and Alloying in Ultrathin Transition-Metal Chalcogenide Nanowires. **ACS Nano**, 10, 2782-2790 (2016).

[40] Yin K, Zhang Y-Y, Zhou Y, Sun L, Chisholm M F, Pantelides S T and Zhou W. Unsupported single-atom-thick copper oxide monolayers. **2D Mater.**, 4, 011001 (2017).

[41] Zhao X, Dan J, Chen J, Ding Z, Zhou W, Loh K P and Pennycook S J. Atom-

[41] by-atom fabrication of monolayer molybdenum membranes. ***Adv. Mater.***, 30, 1707281 (2018).

[42] Sang X, Li X, Zhao W, Dong J, Rouleau C M, Geohegan D B, Ding F, Xiao K and Unocic R R. In situ edge engineering in two-dimensional transition metal dichalcogenides. ***Nat. Commun.***, 9, 2051 (2018).

[43] Zhao X, Qiao J, Chan S M, Li J, Dan J, Ning S, Zhou W, Quek S Y, Pennycook S J and Loh K P. Unveiling Atomic-Scale Moire Features and Atomic Reconstructions in High-Angle Commensurately Twisted Transition Metal Dichalcogenide Homobilayers. ***Nano Lett.***, 21, 3262-3270 (2021).

[44] Clark N, Kelly D J, Zhou M, Zou Y-C, Myung C W, Hopkinson D G, Schran C, Michaelides A, Gorbachev R and Haigh S J. Tracking single adatoms in liquid in a transmission electron microscope. ***Nature***, 609, 942-947 (2022).

[45] Van Winkle M, Dowlatshahi N, Khaloo N, Iyer M, Craig I M, Dhall R, Taniguchi T, Watanabe K and Bediako D K. Engineering interfacial polarization switching in van der Waals multilayers. ***Nat. Nanotechnol.***, DOI: 10.1038/s41565-024-01642-0 (2024).

[46] Meyer J C, Eder F, Kurasch S, Skakalova V, Kotakoski J, Park H J, Roth S, Chuvilin A, Eyhusen S and Benner G. Accurate measurement of electron beam induced displacement cross sections for single-layer graphene. ***Phys. Rev. Lett.***, 108, 196102 (2012).

[47] Pennycook S J and Boatner L A. Chemically sensitive structure-imaging with a scanning-transmission electron-microscope. ***Nature***, 336, 565-567 (1988).

[48] Pennycook S J and Jesson D E. High-resolution z-contrast imaging of crystals. ***Ultramicroscopy***, 37, 14-38 (1991).

[49] Krivanek O L, Chisholm M F, Nicolosi V, Pennycook T J, Corbin G J, Dellby N, Murfitt M F, Own C S, Szilagyi Z S, Oxley M P, Pantelides S T and Pennycook S J. Atom-by-atom structural and chemical analysis by annular dark-field electron microscopy. ***Nature***, 464, 571-574 (2010).

[50] Gong Y, Liu Z, Lupini A R, Shi G, Lin J, Najmaei S, Lin Z, Elias A L, Berkdemir A, You G, Terrones H, Terrones M, Vajtai R, Pantelides S T, Pennycook S J, Lou J, Zhou W and Ajayan P M. Band Gap Engineering and Layer-by-Layer Mapping of Selenium-Doped Molybdenum Disulfide. ***Nano Lett.***, 14, 442-449 (2014).

[51] Li S, Wang Y-P, Ning S, Xu K, Pantelides S T, Zhou W and Lin J. Revealing 3D Ripple Structure and Its Dynamics in Freestanding Monolayer $MoSe_2$ by Single-Frame 2D Atomic Image Reconstruction. ***Nano Lett.***, 23, 1298-1305 (2023).

[52] Lee J, Yang Z, Zhou W, Pennycook S J, Pantelides S T and Chisholm M F. Stabilization of graphene nanopore. ***Proc. Natl. Acad. Sci. USA***, 111, 7522-7526 (2014).

[53] Kresse G and Joubert D. From ultrasoft pseudopotentials to the projector augmented-wave method. ***Phys. Rev. B***, 59, 1758 (1999).

[54] Perdew J P, Burke K and Ernzerhof M. Generalized gradient approximation made simple. ***Phys. Rev. Lett.***, 77, 3865 (1996).

[55] Hafner J. *Ab-initio* simulations of materials using VASP: Density-functional theory and beyond. ***J. Comput. Chem.***, 29, 2044-2078 (2008).

[56] Baroni S, de Gironcoli S, Dal Corso A and Giannozzi P. Phonons and related crystal properties from density-functional perturbation theory. ***Rev. Mod. Phys.***, 73, 515 (2001).

[57] Togo A, Oba F and Tanaka I. First-principles calculations of the ferroelastic transition between rutile-type and $CaCl_2$-type $SiO_2$ at high pressures. ***Phys. Rev. B***, 78, 134106 (2008).